\documentstyle[11pt]{article}
\newcommand{\rf}[1]{(\ref{#1})}
\newcommand{\beq}{\begin{equation}}
\newcommand{\eeq}{\end{equation}}
\newcommand{\bdm}{\begin{displaymath}}
\newcommand{\edm}{\end{displaymath}}
\newcommand{\bea}{\begin{eqnarray}}
\newcommand{\eea}{\end{eqnarray}}
\newcommand{\nn}{\nonumber \\}

\newcommand{\pa}{\partial}

\def\gsim{\raise.3ex\hbox{$>$\kern-.75em\lower1ex\hbox{$\sim$}}}
\def\lsim{\raise.3ex\hbox{$<$\kern-.75em\lower1ex\hbox{$\sim$}}}

\def\vek#1{{\bf #1}}
\begin{document}
\topmargin -1.4cm
\oddsidemargin -0.8cm
\evensidemargin -0.8cm
\title{\Large{Dual strings and magnetohydrodynamics }}

\vspace{0.5cm}

\author{{\sc P. Olesen$^1$ }\\{\sl The Niels Bohr Institute,
University of Copenhagen}\\{\sl Blegdamsvej 17, DK-2100 Copenhagen, Denmark}}
\maketitle
\vfill
\begin{abstract} We investigate whether dual strings could be solutions
of the magnetohydrodynamics equations in the limit of infinite conductivity.
We find that the induction equation is satisfied, and we discuss
the Navier-Stokes equation (without viscosity) with the Lorentz
force included. We argue that the dual string equations (with a
non-universal maximum velocity) should describe the large scale motion
of narrow magnetic flux tubes, because of a large reparametrization (gauge)
invariance of the magnetic and electric string fields. It is shown that
the energy-momentum tensor for the dual string can be reinterpreted as
an energy-momentum tensor for magnetohydrodynamics, provided certain
conditions are satisfied. We also give a brief discussion of the case
when magnetic monopoles are included, and indicate how this can lead to
a non-relativistic "electrohydrodynamics" picture of confinement.

\end{abstract}
\vfill
\begin{flushleft}
NBI-HET-9536\\
hep-th/9509023
\end{flushleft}
\footnoterule
{\small    $^1$E-mail: polesen@nbivax.nbi.dk}
\thispagestyle{empty}
\newpage
\setcounter{page}{1}
A conducting gas/fluid interacting with a magnetic field is described
by magnetohydrodynamics (see ref. \cite{dynamo} for a review of this
subject). A turbulent plasma has a strong tendency to become
extremely intermittent, because the vorticity concentrates in
thin vortex tubes, and the magnetic field concentrates in thin
flux tubes. When one looks at computer simulations \cite{aake} of such a
plasma, the situation looks very "stringy", in the sense of an
intermittent network of "string"-like objects. This suggested to the
author that it would be an interesting question to ask if dual
strings (see e.g. the reviews mentioned in ref. \cite{scherk}) could be
solutions to the magnetohydrodynamics equations.

In the following we shall assume that the conductivity is infinite, and
the viscosity vanishes. In computer simulations one sees [2] that the
flux tubes are formed as parallel magnetic field lines over a long
scale, but then the field lines spread out and the magnetic field
becomes diffuse and weak.  After some distance the field
lines may again join in a flux tube. The diffuse behavior is not
"string"-like, and is therefore from the very beginning not
covered by our considerations. We do therefore not claim that all
features considered in the following are realistic, but we hope that some
of our results are of interest.

In order to have a "stringy" situation, we consider flux tubes which
are closed (or have an infinite length). We also assume that they
move with the velocity of the fluid
at any point. This puts severe restrictions on the dynamics to which our
picture applies. In particular, we expect that the magnetic energy is
roughly of the same order as the kinetic energy of the gas/fluid.

One case where our considerations may be of direct use, is the case
where somehow closed magnetic flux tubes are generated in the early
universe, e.g. as defects in some phase transition. Then these
flux tubes are expected to follow the Nambu-Goto string
equations \cite{scherk}-\cite{cosmic},
at least to the first approximation. On the other hand, these tubes
should also follow the magnetohydrodynamics equations, since such
flux tubes interact with the charged particles. Thus, if these closed
flux tubes are to keep their identity for some time (ultimately they are
expected to decay), the Nambu-Goto string equations $and$ the
magnetohydrodynamics equations should be satisfied. Now the results in
the following indicate that it is indeed possible to satisfy both set of
equations at the same time.

We start by considering a magnetic field in the form of an infinitely
narrow flux tube
\beq
B_i(\vek x,t)=b\int d\sigma {\pa z_i(\sigma,t)\over\pa \sigma}
{}~\delta^3(x-z(\sigma,t))~.
\label{1}
\eeq
The flux tube follows a curve given by $z_i(\sigma,t)$, where $\sigma$ is some
parameter. The constant $b$ has dimension $\sqrt{\rm energy\times\rm length}$.
Since we do not assume that the flux tube has a topological origin, $b$ is
not assumed to have a universal significance. We assume that the string
moves with the same velocity as the fluid velocity ${\bf v}(\vek x,t)$, i.e.
\beq
v_i(\vek x,t)~\delta^3(x-z(\sigma,t))
={\pa z_i(\sigma,t)\over\pa t}~\delta^3(x-z(\sigma,t)),
\label{2}
\eeq
and $\vek B$ is assumed to be perpendicular to $\vek v$. It is also
assumed that the curve described by $\vek z(\sigma,t)$ has no
self-crossings or touchings, so that the tangent is unique.

The non-universal quantity $b$ has the interpretation of the magnetic flux
through a surface perpendicular to $\vek B$, as is seen by integrating
the expression \rf{1} over such a surface,
\beq
\int d^2 x_{\perp}~\vek B(\vek x,t)=b \int d^2 x_{\perp} \int d \vek z~
\delta^3(x-z)=b,
\label{fk}
\eeq
where $d^2 x_{\perp}$ is perpendicular to $\vek B(\vek x,t)$, and hence is
also perpendicular to $d \vek z$. The constancy of $b$ is an expression of the
well-known magnetohydrodynamics conservation of a magnetic flux which follows
the motion of the fluid.

Evidently eq. \rf{1} represents the large scale behaviour of the magnetic
field. It should be mentioned that an identical expression is used in
hydrodynamics and in superfluid helium for the large scale behaviour
of vortex lines (see ref. \cite{nemi} for a recent review). Here the
vorticity $\omega$ is given by
\beq
\omega(\vek x,t)={\rm curl}~\vek v(\vek x,t) ={\rm const.}\int d\sigma
{\pa \vek z(\sigma,t)\over\pa \sigma}~ \delta^3(x-z(\sigma,t)),
\label{hydro}
\eeq
so the resulting velocity is of the Biot-Savart type.

Taking the time derivative of eq. \rf{1} we get
\bea
{\pa B_i(\vek x,t)\over\pa t}&=&b\int d\sigma{\pa^2 z_i\over\pa t\pa \sigma}~
\delta^3(x-z)+b\int d\sigma{\pa z_i\over\pa \sigma}{\pa z_k\over\pa t}~
{\pa\over\pa z_k}\delta^3(x-z)\nn
&=&b\int d\sigma{\pa^2 z_i\over\pa t\pa \sigma}~\delta
^3(x-z)-{\pa (v_k B_i)\over\pa x_k}
\label{3}
\eea
Using that
\beq
(\nabla\times(\vek v\times\vek B))_i={\pa(v_i B_k)\over\pa x_k}
-{\pa(v_k B_i)\over\pa x_k},
\label{4}
\eeq
as well as eq. \rf{2}, we obtain the equation
\beq
{\pa \vek B\over\pa t}=\nabla\times(\vek v\times\vek B),
\label{a}
\eeq
where we used
\beq
{\pa (v_i B_k)\over\pa x_k}=b{\pa\over\pa x_k}\int d\sigma {\pa z_i\over\pa t}
{\pa z_k\over\pa \sigma}~\delta^3(x-z)=-b\int d\sigma {\pa z_i\over\pa t}
{}~{\pa\over\pa \sigma}\delta^3(x-z)=b\int d\sigma {\pa^2 z_i\over\pa \sigma\pa
t}
{}~\delta^3(x-z).
\label{341}
\eeq
Eq. \rf{a} is the well known hydromagnetic form of the induction
equation in the limit of infinite conductivity. Thus the "string"
(i.e. narrow flux tube) \rf{1} satisfies this equation
for $any$ curve $z_i(\sigma,t)$.

The field $\vek B$ should also be divergenceless. It is quite easy to
check that this is the case:
\beq
{\rm div}~\vek B(\vek x,t)=b\int d\sigma {\pa z_i\over\pa \sigma}
{}~{\pa \over\pa x_i}\delta^3(x-z)=-b\int d\sigma{\pa \over\pa \sigma}
\delta^3(x-z)=0.
\label{5}
\eeq
Here we have used that the "string" is assumed to be closed, so that the
contributions from the boundaries exactly cancel in eq. \rf{5}.

In the limit of infinite conductivity the only existing electric field
is the one induced by $\vek B$, namely $c~\vek E=-\vek v\times\vek B$, where
$c$ is the velocity of light, which cancels out in eq. \rf{a}. Now one
of the main points of this paper is that the
electric and magnetic fields can be unified in the tensor field
\beq
F_{\mu\nu}(\vek x,t)=b \int d\sigma d\tau \left({\pa z_{\mu}\over\pa \sigma}
{\pa z_{\nu}\over\pa \tau}-{\pa z_{\mu}\over\pa \tau}
{\pa z_{\nu}\over\pa \sigma}\right)
\delta^4(x-z(\sigma,\tau)).
\label{6}
\eeq
Here $\tau$ is an arbitrary "time" parameter, $x_{\mu}$ in the delta
function is the four vector $(x_0=t,\vek x)$, and $z_{\mu}$ is the
four vector $(z_0,\vek z)$. In the special parametrization $z_0(\sigma,\tau)$
=$\tau$, we have
\beq
F_{0i}=-B_i,~~F_{ij}=c~  \epsilon_{ijk} E_k,
\label{7}
\eeq
Eqs. \rf{a} and \rf{5} can also be unified in the single equation
\beq
{\pa F_{\mu\nu}(x)\over\pa x_{\mu}}=0.
\label{ny}
\eeq
Here we used the boundary condition that everything vanishes at $\tau=
+\infty$ and $\tau=-\infty$. The crucial point is now that
the expression \rf{6} is reparametrization invariant, i.e.
$F_{\mu\nu}$ is the same no matter which set of parameters $(\sigma,\tau)$
we use: A reparametrization, where $\sigma$ and $\tau$ are replaced by
$\tilde{\sigma}(\sigma,\tau)$ and $\tilde{\tau}(\sigma,\tau)$, respectively,
leaves $F_{\mu\nu}$ invariant, and hence the physics does not
depend on the parameters we use. The parametrization $z_0(\sigma,\tau)=\tau$
used in the beginning is thus very special. In the general case
$z_{\mu}(\sigma,\tau)$ describes a surface in space and time,
to be identified with the "world sheet" of the string associated with
the (infinitely) narrow flux tube. If the time $t$ is replaced by an
arbitrary "time" parameter $\tau$, we thus have a very large "gauge" freedom,
consisting of the freedom of choosing parameters $\sigma$ and $\tau$
describing the same surface. Under very general conditions we can
restrict this freedom by choosing an orthonormal system (see
ref. \cite{scherk}).
Now we have already taken the velocity perpendicular to the flux tube, i.e.
\beq
{\pa z_i\over\pa t}~{\pa z_i\over\pa \sigma}=0.
\label{8}
\eeq
In order to have orthonormality we should further normalize the velocity
and the tangent vectors according to
\beq
\left({\pa \vek z\over\pa \sigma}\right)^2+
\frac{1}{v^2_0}\left({\pa \vek z\over\pa t}\right)^2=1.
\label{9}
\eeq
Here $v_0$ is a velocity introduced for dimensional reasons. In string theory
the corresponding velocity is universal and equal to the velocity of light.
Here we do not assume that $v_0$ is in any way universal. From eq. \rf{9}
it follows that $v_0$ is the maximum (transverse) velocity the flux tube
considered can have.

It is now easy to see that the motion of the string is fixed to be given by
the usual harmonic equation of motion for $z_i(\sigma,t)$. To see this,
differentiate the normalization condition \rf{9} with respect to $\sigma$,
\beq
{\pa^2 z_i\over\pa t\pa \sigma}{\pa z_i\over\pa t}
+v^2_0{\pa^2 z_i\over\pa \sigma^2}{\pa z_i\over\pa \sigma}=0.
\label{13}
\eeq
{}From the orthogonality condition \rf{8} we obtain by differentiation
with respect to $t$
\beq
{\pa^2 z_i\over\pa t^2}{\pa z_i\over\pa \sigma}+{\pa z_i\over\pa t}
{\pa^2 z_i\over\pa t\pa \sigma}=0.
\label{14}
\eeq
Using this in eq. \rf{13} we get
\beq
v^2_0{\pa^2 z_i\over\pa \sigma^2}{\pa z_i\over\pa \sigma}=
{\pa^2 z_i\over\pa t^2}{\pa z_i\over\pa \sigma},
\label{15}
\eeq
which has the solution
\beq
{\pa^2 z_i\over\pa \sigma^2}=\frac{1}{v^2_0}{\pa^2 z_i\over\pa t^2}+
\alpha(\vek z^2,t) {\pa z_i\over\pa t},
\label{16}
\eeq
where $\alpha$ is an arbitrary function. In the following we shall take
$\alpha=0$. The function $z_i(\sigma,t)$ thus satisfies the harmonic equation
together with the orthonormality conditions \rf{8} and \rf{9}. The solution
to these non-linear equations is known from strings. The harmonic equation
for $z_i$ has the solution
\beq
z_i(\sigma,t)=q_i+p_i t+i\sum_{n=-\infty}^{+\infty} \frac{1}{n}
\exp\left(-inv_0\frac{t}{l}\right)\left(\alpha_{n,i}
\cos n\frac{\sigma}{l}+\beta_{n,i} \sin n\frac{\sigma}{l}\right),
\label{17}
\eeq
where $l$ is some lenght scale, which does not have to be universal.
The non-linear orthonormality conditions are solved by the conditions
\beq
L_n(\alpha)=L_n(\beta)=0,
\label{18}
\eeq
where
\beq
L_n(x)=\sum x_{n-m}~x_m,~ x=\alpha~~{\rm or}~~\beta.
\label{19}
\eeq
Here the sum over $m$ runs from $-\infty$ to $+\infty$,and a summation
over the space indices is understood. Also, we have defined $\alpha_{0,i}
=\beta_{0,i}$ as beeing proportional to the constant $p_i$. We refer
to the literature for further discussion of the solution \cite{scherk}.

We thus have the situation that the non-linear eqs. of motion can be solved
in principle, and the motion of the string can be followed. In practice,
however, the solution given in eqs. \rf{17}-\rf{19} is somewhat formal,
and it should presumably be replaced by numerical solutions, where one
starts from an initial curve and study the subsequent development.

We have seen that the fundamental magnetohydrodynamics equation \rf{a} is
satisfied by the string ansatz \rf{1}. Now one of the central points of
this paper is that if we make the $ansatz$ \rf{1}, then the existence of
a more general underlying physical theory (Maxwell's) implies that
the dynamics is really described by the field tensor $F_{\mu\nu}$ \rf{6}.
This implies a very large freedom of
parametrization, which can be
restricted to be orthonormal (see ref. [3]) like in eqs. \rf{8} and
\rf{9}. Thus, any velocity field $\vek v(\vek x,t)$, which is
consistent with the ansatz \rf{1} or \rf{6}, must necessarily satisfy
the harmonic eq. \rf{16}. Thus it follows that {\sl either a pressure and a
density field can be found such that the corresponding solution of the
the Navier-Stokes equation also has the properties \rf{8}, \rf{9}, and \rf{16},
or the ansatz \rf{1} or \rf{6} is not consistent with
magnetohydrodynamics.} Encouraged by the fact that the ansatz \rf{1}
or \rf{6} satisfies the induction eq. \rf{a}, and by the result of
computer calculations \cite{aake}, we shall tentatively assume
that the first possibility is realized. We therefore consider the Navier-Stokes
equation, or rather the Euler equation including the Lorentz-force (since
there is no viscosity term $\mu\nabla^2\vek v$), i.e.
\beq
{\pa \rho v_i\over\pa t}+{\pa \rho v_k v_i\over\pa x_k}
=-{\pa P\over\pa x_i}+((\nabla\times \vek B)\times \vek B)_i
=-{\pa\over\pa x_i}(P+\frac{1}{2}\vek B^2)+B_k{\pa\over\pa x_k} B_i.
\label{ns}
\eeq
Here $P(\vek x,t)$ is the local pressure, and $\rho(\vek x,t)$ is the
local mass density. For simplicity, we consider the non-relativistic
version of the Navier-Stokes equation. Since $z_i(\sigma,t)$ in principle
is known as a solution of the string eqs. \rf{8}, \rf{9}, and \rf{16} with
$\alpha=0$, this amounts to finding out the conditions
imposed by equation \rf{ns} on the pressure $P$ and
the density $\rho$. This is a non-trivial problem, since the
gas/liquid may or may not penetrate the flux tube. We shall discuss
this problem later.

It should be noticed that the Navier-Stokes equation couples the magnetic
field and the velocity dynamically, so the field $\vek B$ reacts back on
the velocity field. Thus our previous statement that the
field $\vek B$ is supposed to move with the fluid in the induction
equation \rf{a} is to be understood in the sense that $\vek v$ is a
functional of $\vek B$.

At this stage the reader may wonder why one cannot use the arguments
leading to the Nambu-Goto equations \rf{8}, \rf{9}, and \rf{16} in the
case of the vorticity \rf{hydro}. The difference is that e.g. in the
hydrodynamics case the fundamental field is the velocity, and
$\omega$ is a quantity {\sl derived} from $\vek v$. There is no physical
point in considering a "unified" tensor $(\omega,\vek v\times\vek\omega)$,
and claim that this is somehow the fundamental tensor in hydrodynamics.
This is in contrast to the Maxwell equations, which unifies $\vek E$
and $\vek B$ physically. Therefore, it has a physical meaning to
exploit the large gauge invariance in $F_{\mu\nu}$, which is due
to the reparametrization invariance, in much the same way that
ordinary gauge invariance is often used to choose a simple gauge
where the calculations are done in the most transparent way.

We shall now compare with the Navier-Stokes eq. \rf{ns}, which can be derived
from the momentum flux density tensor (see Landau and Lifshitz \cite{dynamo},
p. 215)
\beq
\Pi_{ik}=\rho v_i v_k-B_i B_k+(P+\frac{1}{2}\vek B^2)\delta_{ik},
\label{-1}
\eeq
with
\beq
{\pa \rho v_i\over\pa t}=-{\pa \Pi_{ik}\over\pa x_k}.
\label{-2}
\eeq
These equations are of course non-relativistic.
We would like to compare $\Pi_{ik}$ with the energy momentum tensor for the
string. In order to do this, we need first to discuss the quadratic terms
in $\vek B$ in eq. \rf{-1}. In computing the quadratic terms from the
ansatz \rf{1} we encounter divergencies, due to the product of two delta
functions. We therefore need a regulator, which we can take to be Gaussian,
\beq
\delta^3(x)\approx \frac{1}{\pi^{\frac{3}{2}}l^3_\perp}~
\exp\left(-\frac{\vek x^2}{l^2_\perp}\right),
\label{-3}
\eeq
where $l_\perp$ is the transverse dimension of a flux tube replacing the
string, and where $l_\perp$ is assumed to be small relative to all other
distances, of course. For simplicity we have regulated the longitudinal
direction with the same regulator as the transverse directions. Regulators
of the type \rf{-3} have been used in the study of superfluid
helium \cite{nemi}.

Let us now consider one of the quadratic terms on the right hand side of eq.
\rf{-1}. Formally we have
\beq
B_kB_i=b^2\int d\sigma d\sigma'
z'_k(\sigma,t)z'_i(\sigma',t)~\delta^3(x-z(\sigma,t))\delta^3(x-z(\sigma',t)),
\label{-4}
\eeq
where the prime on $\vek z$ denotes the derivative with respect to the first
argument. If we use this expression, we have an infinity.
If we insert the regulator \rf{-3}, we can use that the integrals have support
only when $\vek x=\vek z(\sigma,t)=\vek z(\sigma',t)$, to within the
distance $l_\perp$. Thus, to the first approximation,  we can replace
one of the delta functions by its value at zero argument. If $\sigma_0$ is the
parameter which satisfies $\vek x=\vek z(\sigma_0,t)$, then the requirement
that $\vek x\approx \vek z(\sigma,t)$ to an accuracy of order $l_{\perp}$, is
satisfied for $|\sigma_0-\sigma|\sim l_{\perp}\ll l$, as is seen by expanding
the solution \rf{17} around $\sigma_0$. Thus the effective range of the
$\sigma$-integration is of order $l_{\perp}$. Up to a numerical constant we
thus obtain
\beq
B_kB_i\approx \frac{b^2}{\pi^{\frac{3}{2}}l^2_\perp}
\int d\sigma z'_k(\sigma,t)z'_i(\sigma,t)~\delta^3(x-z(\sigma,t)).
\label{-xx}
\eeq
The moral lesson of this exercise is that we $cannot$ accept the ansatz
\rf{1} as an exact string equation, but we must always think of the string as
a narrow flux tube.

Now we shall show that the magnetohydrodynamics equations based on the
ansatz \rf{1} and discussed above, can be obtained by a reinterpretation of the
string energy-momentum tensor. The latter has been used often in the study
of cosmic strings (see ref. \cite{cosmic}), and is given by
\beq
T^{\rm string}_{\mu\nu}=\frac{b^2}{\pi^{\frac{3}{2}}l^2_\perp}
{}~\int d \sigma d \tau \left(\frac{\dot{z_{\mu}}
\dot{z_{\nu}}}{v_0^2}- z'_{\mu} z'_{\nu}\right)~\delta^4(x-z(\sigma,\tau)),
\label{-5}
\eeq
where we use four-vector notation, and where the dot indicate the derivative
with respect to $\tau$. The constant in front is introduced for reasons to
be made clear in the following. Like in eq. \rf{6}, $T^{\rm string}_{\mu\nu}$
is a priori defined in terms of the arbitrary parameters $\sigma$ and $\tau$,
but in the expression \rf{-5} we have already used the orthonormal "gauge"
for simplicity (see the string literature \cite{scherk} and \cite{cosmic}
for a discussion of this point). It is easy to see from the harmonic eq.
\rf{16} that there is conservation of energy and momentum flow,
\beq
{\pa T^{\rm string}_{\mu\nu}\over\pa x_{\mu}}=0.
\label{-6}
\eeq
The boundary terms disappear because the string is closed and because we
assume that everything disapppear at infinite times. Now from eq. \rf{-5}
we have, by doing the $\tau$-integral
\beq
T^{\rm string}_{0i}=\frac{b^2}{\pi^{\frac{3}{2}}l^2_\perp v^2_0}
\int d \sigma \dot{z_i} \delta^3(x-z(\sigma,t))
\equiv \rho v_i,
\label{-7}
\eeq
where we have defined
\beq
\rho(x)=\frac{b^2}{\pi^{\frac{3}{2}}l^2_\perp v^2_0} \int d \sigma
\delta^3(x-z(\sigma,t)).
\label{-8}
\eeq
It is easy to see that $\rho$ satisfies the continuity equation, and
that it has the dimension of a mass density. We also have
\beq
T^{\rm string}_{ki}=\rho v_k v_i-\frac{b^2}{\pi^{\frac{3}{2}}l^2_\perp}
\int d \sigma z'_kz'_i~\delta^3(x-z(\sigma,t))=\rho v_k v_i-B_kB_i.
\label{-9}
\eeq
Here we used eq. \rf{-xx} in the last step.
It is now seen that the magnetohydrodynamics equation in the form \rf{-2} is
equivalent to the string energy-momentum conservation \rf{-6} with $\nu=i$,
provided the density of the fluid inside the string is identified as in
eq. \rf{-8}, and provided we take
\beq
P+\frac{1}{2} \vek B^2={\rm const}.
\label{-10}
\eeq
This condition means that the space-space components of two tensors \rf{-1}
and \rf{-5} differ by a constant only. Such a difference is without
dynamical importance.

It is interesting that the condition \rf{-10} is the same as the variation
of the pressure in the exact hydromagnetodynamic wave discussed by
Landau and Lifshitz (\cite{dynamo}-see p. 221). Eq. \rf{-10} tells
us that the hydrodynamic pressure should have a dip inside the string (flux
tube).

Assuming the ansatz defined by eqs. \rf{1},\rf{2},\rf{-8}, and \rf{-10},
we therefore have derived the result that
the {\sl dual string energy-mometum tensor can be reinterpreted as the energy-
momentum tensor of magnetohydrodynamics in the limit of infinite conductivity}.
The classical motion of the dual string can thus be understood as a
motion in a fluid under the influence of a magnetic field generated by
the string itself as in eq. \rf{6}.
In this connection it should be remembered that the string $cannot$ be
considered as an ideal zero-width string, but must be given a small
transverse dimension. Of course, the same is true for cosmic strings,
where eq. \rf{-5} is also used \cite{cosmic}, where the global properties are
well described by this string energy-momentum tensor.

A simple, but not very interesting, closed string solution is given by
\beq
\vek z(\sigma,t)=l~\cos~\omega t~\left(\cos \frac{\sigma}{l},\sin
\frac{\sigma}{l} \right),
\label{-11}
\eeq
where $l$ is the lenght scale introduced previously, and where the
maximum velocity is given by $v_0=l\omega$. For $0 \leq~\omega t~\leq \pi$
/2, this solution represents a circular flux tube with decreasing radius,
corresponding to a fluid which moves towards the origin. It is more interesting
that many results are known about dual strings (see refs.
\cite{scherk}-\cite{cosmic}, where further references can be found), so we
hope that the re-interpretation mentioned above can be of use in
magnetohydrodynamics. Conversely, in cosmology it may be of interest that
magnetic strings follow the magnetohydrodynamics way of life.

We end this discussion by emphasizing that string-like (or flux tube) solutions
have previously been considered by many authors, as is clear from the
literature
quoted in ref. \cite{dynamo}. A recent review by Semenov is given in
\cite{sun}.
He also derives harmonic equations, but they differ from ours, since in his
case the term $\vek z''$ is multiplied by the density $\rho$, and
orthonormality
is not used. His solution is more general than ours, since the condition
\rf{-10} is not imposed. In general, the energy-momentum tensor corresponding
to Semenov's equations $cannot$ be reinterpreted in terms of dual strings. This
shows that there are other string-like solutions of the magnetohydrodynamics
equations than the $dual$ type considered in the present paper.

So far we have considered closed strings. It is, however, quite easy to add
monopoles. The induction equation \rf{a} is then changed to
\beq
{\pa \vek B\over\pa t}=\nabla\times(\vek v\times\vek B)-\vek j_{\rm magnetic},
\label{-12}
\eeq
where the magnetic current is given by
\beq
\vek j_{\rm magnetic}=b~\dot{\vek z}(l,t)~\delta^3(x-z(l,t))-b~\dot{\vek
z}(0,t)
{}~\delta^3(x-z(0,t)).
\label{-13}
\eeq
Here we inserted the ansatz \rf{1}. The divergence equation \rf{5} is of
course also changed,
\beq
{\rm div}~\vek B(\vek x,t)=-b~\delta^3(x-z(l,t))+b~\delta(x-z(0,t)).
\label{-14}
\eeq
The constant $b$ can be fixed by Dirac's quantization condition.
Thus the generalised magnetohydrodynamics equations have magnetic strings
between the magnetic monopoles, which may therefore be confined. Of course, we
do not know if there are non-confining solutions which are more
favoured energetically. For the sake of the argument, let us assume that these
strings are favoured. Then we have a pedagogically simple picture of quark
confinement: We know that Maxwell's equations are invariant under dual
transformations. Hence, we can imagine a situation where we have a
"universe" (vacuum) filled with magnetic monopoles, with a "dual Ohm's law",
\beq
\vek j_{\rm magnetic}=\sigma_{\rm magnetic}~\vek B,
\label{sveske}
\eeq
where $\sigma_{\rm magnetic}$ is the conductivity of the monopoles, and
where eq. \rf{sveske} is valid in the rest frame.  Assuming now that
the magnetic conductivity is infinite, we would have "electrohydrodynamics"
which is dual to the usual magnetohydrodynamics. Thus the $\vek B$-field
vanishes in the rest frame of the fluid. Thus,
for example, instead of eq. \rf{-12} we would have
\beq
{\pa \vek E\over\pa t}=\nabla\times(\vek v\times\vek E)-\vek j_{\rm electric},
\label{-15}
\eeq
since the only magnetic field is given by $c~\vek B=\vek v\times\vek E$ in the
moving frame. Eq. \rf{-15} is therefore one of the Maxwell equations. The
electric current has an expression analogous to
eq. \rf{-13}. It should also be assumed that the "dual displacement current"
$\pa \vek B$/$\pa t$ vanishes. The string solution of eq. \rf{-15} is thus
an electric string,
\beq
\vek E=b'\int d\sigma {\pa \vek z(\sigma,t)\over\pa \sigma}~
\delta^3(x-z(\sigma,t)),
\label{kuk}
\eeq
where $b'$ is a constant. Here the field $\vek z(\sigma,t)$ follow the
ortonormal dual string equations. There is an equation analogous to eq.
\rf{-14}
for ${\rm div}~\vek E$. The electric charge, which is related to $b'$, is
thus confined.

This scenario of quark confinement is somewhat similar to the dual
superconductor picture \cite{conf}. It is perhaps easier to imagine
electrohydrodynamics than it is to imagine a dual superconductor, because the
former is a simple consequence of the duality inherent in the Maxwell
equations.
On the other hand, it is not clear if our simple picture can be generalized
relativistically, and hence it can only be regarded as some low energy
approximation. In any case, a crucial ingredient in both cases is that
conductivity should be very large and monopoles are needed in the vacuum.
In QCD it may be possible to distinguish the two cases in lattice gauge
calculations, e.g. by checking whether the dual London equations are satisfied
\cite{london}. In electrohydrodynamics these dual superconductor equations are
not expected to hold.

\vskip0.4truecm
I thank \AA ke Nordlund for an illustrative demonstration of his amazing
computer simulations of magnetic flux tubes, and I thank Axel Brandenburg
for communicating his knowledge and insight in magnetohydrodynamics
in an almost infinite number of conversations. Thanks are also due to
Mark Hindmarsh for pointing out the existence of ref. \cite{sun}.

\end{document}